\documentclass[twocolumn, showpacs, prb]{revtex4}

\usepackage{graphicx}

\usepackage{dcolumn}

\usepackage{bm}

\usepackage{color}

\begin{document}

\title{Composite fermion valley polarization energies: Evidence for particle-hole asymmetry}

\date{\today}

\author{Medini Padmanabhan}

\author{T.\ Gokmen}

\author{M.\ Shayegan}

\affiliation{Department of Electrical Engineering, Princeton
University, Princeton, NJ 08544}

\begin{abstract}

In an ideal two-component two-dimensional electron system,
particle-hole symmetry dictates that the fractional quantum Hall
states around $\nu = 1/2$ are equivalent to those around $\nu =
3/2$. We demonstrate that composite fermions (CFs) around $\nu =
1/2$ in AlAs possess a valley degree of freedom like their
counterparts around $\nu = 3/2$. However, focusing on $\nu = 2/3$
and 4/3, we find that the energy needed to completely valley
polarize the CFs around $\nu = 1/2$ is considerably smaller than the
corresponding value for CFs around $\nu = 3/2$ thus betraying a
particle-hole symmetry breaking.

\end{abstract}


\maketitle

In the case of two-dimensional electron systems (2DESs) with no
discrete degrees of freedom such as spin or valley, the principle of
particle-hole symmetry dictates that the fractional quantum Hall
(FQH) state formed at a Landau level (LL) filling factor of $\nu$ is
equivalent to the state formed at ($1-\nu$). In the presence of a
discrete degree of freedom (e.g. spin in the case of GaAs 2DES), the
symmetry relates the states at $\nu$ and ($2-\nu$). Although the
principle of particle-hole symmetry is theoretically sound, its
applicability in real experimental systems is not well studied
\cite{footnote1}. Spin-polarization studies in GaAs have hinted
towards differences between FQH states formed around $\nu$ = 1/2 and
$(2-1/2) = 3/2$
\cite{engelPRB92,duPRL95,duPRB97,kukushkinPRL99,sternPRB04}.
However, while comparing experimental results to theoretical
calculations, it is commonly assumed that particle-hole symmetry
holds \cite{duPRL95,parkPRL98}.

In this paper, we provide detailed and quantitative experimental
evidence proving that particle-hole symmetry is violated in 2DESs.
The quantity we study is the valley-polarization energy of composite
fermions (CFs) in AlAs 2DESs, where electrons occupy two conduction
band minima (valleys). The occupation of these valleys can be
controlled via the application of in-plane strain. Here, we
demonstrate that, similar to their counterparts around $\nu$ = 3/2,
CFs around $\nu = 1/2$ also have their own valley degree of freedom.
We valley polarize these CFs by applying an in-plane uniaxial
strain. The energies required to completely valley polarize the CFs
around $\nu = 1/2$ and 3/2, normalized to the Coulomb energy, should
be identical if particle-hole symmetry prevailed. Surprisingly, we
find that it takes much less energy to completely valley polarize
the CFs around $\nu = 1/2$ compared to the CFs around 3/2. In
particular, we investigate the valley polarization of FQH states at
$\nu = 2/3$ and 4/3 in a wide range of 2D electron densities ($n$),
and conclude that particle-hole symmetry is violated in our system.

\begin{figure}
\includegraphics[scale=1]{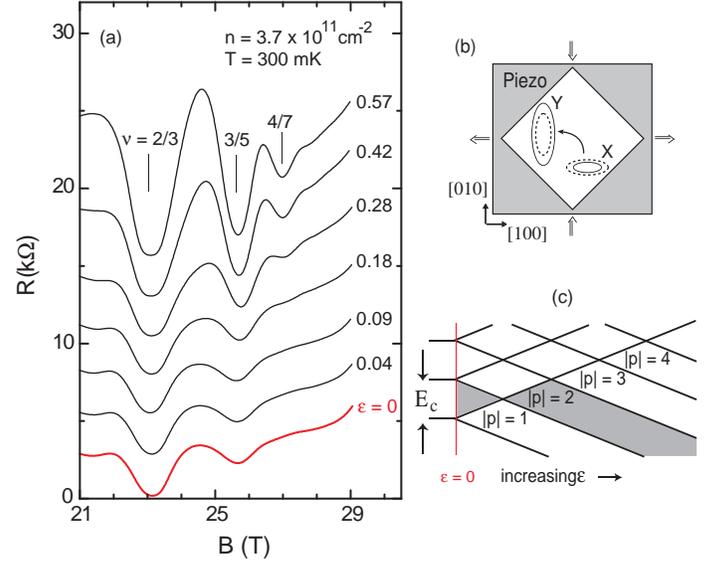}
\caption{(a) Magnetoresistance traces taken at different values of
strain (indicated in units of $10^{-4}$) showing FQH states in the
lowest Landau level. The traces are offset vertically for clarity.
(b) Schematic showing the strain-induced
transfer of electrons from one valley to another.(c)
Simple fan diagram showing the response of composite fermion Landau
levels to strain.}
\end{figure}

Our sample is a 12 nm-wide AlAs quantum well, grown by molecular
beam epitaxy on a semi-insulating (001) GaAs substrate and
modulation doped with Si \cite{shayeganPhysicaB06}. We defined a
Hall bar along the [110] crystal direction with standard
photolithography and made GeAuNi contacts. We also deposited
metallic front and back gates, which allowed us to control $n$. Here
we report data from three separate cooldowns. We made measurements
in a 20 mK base temperature dilution refrigerator and a 300 mK
$^3$He system using standard lock-in techniques.

The 2D electrons in our samples occupy two conduction band valleys
with elliptical Fermi contours \cite{shayeganPhysicaB06}. We can
controllably lift the valley degeneracy by applying uniaxial strain
in the plane of the sample. Experimentally, we achieve this by
gluing the sample to a piezoelectric (piezo) stack which expands in
one direction (and contracts in the perpendicular direction) when an
external voltage is applied. A schematic is shown in Fig.\ 1(b). The
single-particle energy splitting between the two in-plane valleys is
given by $E_{v} = \epsilon E_{2}$; $\epsilon =
\epsilon_{[100]}-\epsilon_{[010]}$ where $\epsilon_{[100]}$ and
$\epsilon_{[010]}$ are the strains along the [100] and [010] crystal
directions \cite{shayeganPhysicaB06} and $E_{2}$ is the deformation
potential, which has a band value of 5.8 eV in AlAs. In
our system, for the density range under study, the Zeeman energy
($E_{Z}$) of the electrons is comparable to their cyclotron energy
\cite{shayeganPhysicaB06}. Since the FQH states around $\nu = 1/2$ and
3/2 are formed in the lowest LLs of the two occupied valleys, the spin degree of freedom does
not play a role in the results reported below.

In Fig. 1(a), we show longitudinal resistance ($R$) vs.
perpendicular magnetic field ($B$) traces around $\nu = 1/2$ taken
at different values of $\epsilon$ for a fixed density. At high
values of $\epsilon$, when the CF system is effectively
single-valley, FQH states up to $\nu = 4/7$ are clearly visible. As
we approach the condition of $\epsilon = 0$, $all$ the FQH states
become weak. At first glance, this is very different from the
behavior reported for states around $\nu = 3/2$
\cite{bishopPRL07,padmanabhanPRB09}. In the $\nu = 3/2$ case, the
resistance minima at various fillings get weaker and stronger as a
function of $\epsilon$, a behavior which finds a ready explanation
in terms of the LL energy fan diagram for the CFs (Fig. 1(c))
\cite{bishopPRL07,padmanabhanPRB09}. Note that within the framework
of the CF theory \cite{jainPRL89,halperinPRB93},  FQHE of electrons
is understood as the integer QHE of CFs. Every electronic fractional
filling factor $\nu$ has a CF integer counterpart $p$, related by
$\nu = \frac{p}{2p + 1}$ \cite{CFbook}.

\begin{figure}
\includegraphics[scale=1]{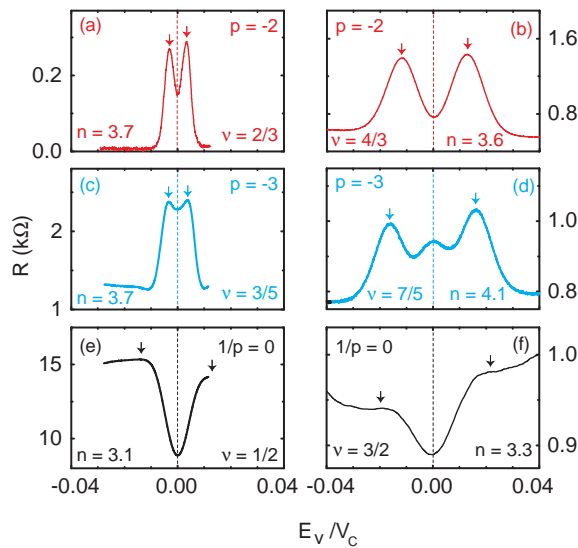}
\caption{ Piezoresistance ($R$ vs. $\epsilon$) traces taken at $T$ =
300 mK and at fixed values of filling factor around $\nu = 1/2$
(left panels) and $\nu = 3/2$ (right panels). The $x$-axis is
expressed as the strain-induced valley splitting ($E_{v}$) measured
in units of the Coulomb energy ($V_{C}$).The values of $\nu$ and $p$
are indicated. The corresponding values of $n$ are also given in
units of $10^{11}$ cm$^{-2}$. The values of $E_{v,pol}$ are
indicated by vertical arrows.}
\end{figure}

We now present data where fixed the value of $\nu$ and obtained
piezoresistance traces. In Fig.\ 2 we compare the response of the
system to $\epsilon$ at and around $\nu = 3/2$ and $\nu = 1/2$. The
x-axis is the valley splitting energy ($E_{v} = \epsilon E_{2}$)
expressed in terms of the Coulomb energy, $V_{C} = e^{2}/4 \pi
\kappa \epsilon_{0}\textit l_{B}$ where $\kappa = 10$ is the
dielectric constant of AlAs and $\textit l_{B}=\sqrt{\hbar/eB}$ is
the magnetic length. The data are taken at comparable values of $n$,
meaning a variation by a factor of about three in $B$. The
normalization of the x-axis ensures that the variation in the values
of $B$ does not affect the energy scales since all relevant energies
are expected to scale with $\sqrt{B}$.

Figures 2(a) and 2(b) show the piezoresistance traces taken at $\nu
= 2/3$ and 4/3 respectively. Both these filling factors correspond
to $p = -2$ of CFs and are therefore expected to behave similarly.
According to the fan diagram in Fig. 1(c), the energy gap ($\Delta$)
at $p = -2$ ($\nu = 2/3,4/3$) is finite at $\epsilon = 0$. As
$\epsilon$ is increased, the CF LLs undergo an energetic coincidence
which results in a vanishing value of $\Delta$. Beyond this
coincidence, the system is completely valley polarized. The energy
at which the coincidence occurs is thus the valley polarization
energy, $E_{v,pol}$. Upon further increasing the strain, the gap
increases, finally saturating at a value equal to the cyclotron
energy ($E_{c}$) of the CFs. This variation in $\Delta$ as a
function of $\epsilon$ affects the value of $R$ at $p = -2$. At any
given temperature ($T$), a small value of gap corresponds to a high value
of $R$ and $vice$ $versa$. Hence, as a function of $\epsilon$, the
resistance at $p = -2$ is expected to oscillate, following the fan
diagram in Fig.\ 1(c). The piezoresistance trace taken at $\nu =
4/3$ (Fig.\ 2(b)) is consistent with this observation. The trace at
$\nu = 2/3$ (Fig.\ 2(a)) qualitatively follows the same trend,
although there is a major quantitative difference. The two peaks
signifying $E_{v,pol}$ on either side of $\epsilon = 0$ are much
closer to each other compared to $\nu = 4/3$. We conclude that the
value of $E_{v,pol}$ for $\nu = 2/3$ is considerably smaller than
for $\nu = 4/3$.

Data for other FQH states around $\nu = 3/2$ and 1/2 lead to a
similar conclusion. For example, in Figs. 2(c) and (d), we show
traces taken at $\nu = 3/5$ and 7/5 respectively. Although the $\nu
= 7/5$ ($p = -3$) trace appears to follow the fan diagram in Fig.
1(c) quite well, $\nu = 3/5$ does not. We believe the value of
$E_{v,pol}$ for $\nu = 3/5$ is so small that, given the finite width
of the piezoresistance peaks, the central peak is not resolved.

Piezoresistance traces taken at $\nu = 1/2$ and 3/2, presented in
Figs. 2(e) and 2(f), show qualitatively similar features. The
resistance in both cases exhibits a minimum when the two valleys are
balanced, increases as the valley degeneracy is broken, and
saturates once the CFs are fully valley polarized. The arrows in
Figs.\ 2(e) and 2(f) denote the upper limits of the estimated valley
polarization energies \cite{padmanabhanPRB09}. Note that the
separation between the arrows is smaller for the case of $\nu = 1/2$
compared to $\nu = 3/2$. In general, the values of $E_{v,pol}$ for
CFs around $\nu = 1/2$ are so small that the response of the system
to $\epsilon$ is contained in a very narrow range of $E_{v}$. This
is why to the first order, all fractions around $\nu = 1/2$ appear
to become weak near $\epsilon = 0$ (Fig. 1(a)).

\begin{figure}
\includegraphics[scale=1]{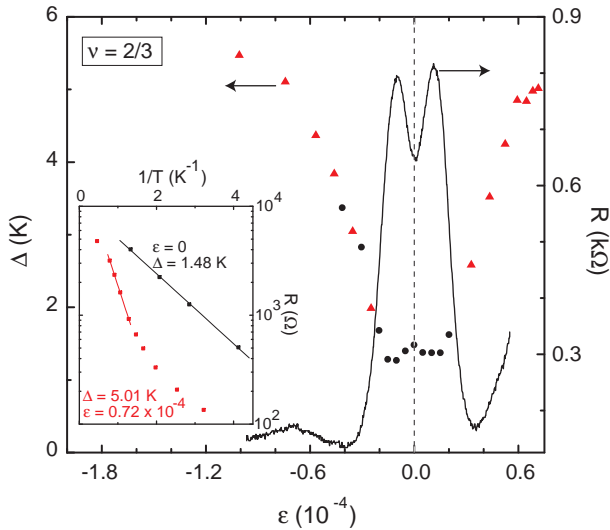}
\caption{Piezoresistance (right y-axis) and energy gaps (left
y-axis) at $\nu = 2/3$. The piezoresistance trace was taken at $B =
16.6$ T ($n = 2.67 \times 10^{11}$ cm$^{-2}$) at $T = 250$ mK. The
gaps shown as (black) circles are obtained for the same $B$ from the
same cooldown. The (red) triangles are gaps obtained from a separate
cooldown at a comparable value of $B = 15.0$ T ($n = 2.41 \times
10^{11}$ cm$^{-2}$). Inset: Arrhenius plots for two extreme values
of $\epsilon$.}
\end{figure}

To further contrast the valley polarization energies for the FQH
states around $\nu = 1/2$ and 3/2, we present data taken at $\nu =
2/3$ in greater detail and compare them to $\nu = 4/3$. Figure 3
shows a piezoresistance trace at $\nu = 2/3$ along with energy gap
measurements. Note that the small dip in the piezoresistance at
$\epsilon = 0$ corresponds to the small increase in the gap
\cite{footnote2}. Also the piezoresistance maxima correspond to the
minima in the gap, which lends support to our determination of the
polarization strain from the piezoresistance measurement. The gap
measurements were done in two different cooldowns because of the
difference in the optimum temperature ranges needed. The Arrhenius
plots for two extreme values of $\epsilon$ are shown in the inset to
Fig.\ 3.

\begin{figure}
\includegraphics[scale=1]{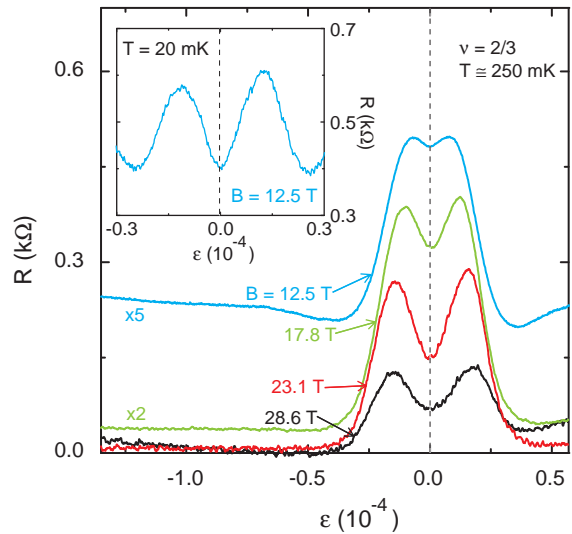}
\caption{Piezoresistance at $T \cong$ 250 mK for $\nu = 2/3$ at
different values of $B$ (corresponding $n$ are 2.0, 2.9, 3.7 and
4.6$\times 10^{11}$cm$^{-2}$). The inset illustrates that the peaks
are better resolved at low temperatures for smaller $B$. The traces
at $B = 12.5$ T and $B = 17.8$ T should be multiplied by the factors
shown to obtain the actual values of resistance.}
\end{figure}

In Fig.\ 4 we show the evolution of the $\nu = 2/3$ piezoresistance
as a function of $n$. The data shown in the main panel were taken at
$T \cong 250$ mK. Note that, as the value of $n$ decreases, the peaks
get closer to each other and almost merge into one at the lowest
density of $n = 2.0 \times 10^{11}$ cm$^{-2}$ ($\nu = 2/3$ at $B$ =
12.5 T). However, for the same $n$, the resolution between the peaks
is considerably improved when temperature is lowered as shown in the
inset. We note that at higher values of $n$, the high temperature is
necessary for the observation of the peaks because, at low values of
$T$, the resistance minimum at $\nu = 2/3$ is too strong and flat to
show any features as a function of $\epsilon$. At intermediate
values of $n$, where peaks are observable at both high and low
temperatures we observe no dependence of the peak positions on $T$,
suggesting that $E_{v,pol}$ is a $T$-independent quantity.

\begin{figure}[b]
\includegraphics[scale=1]{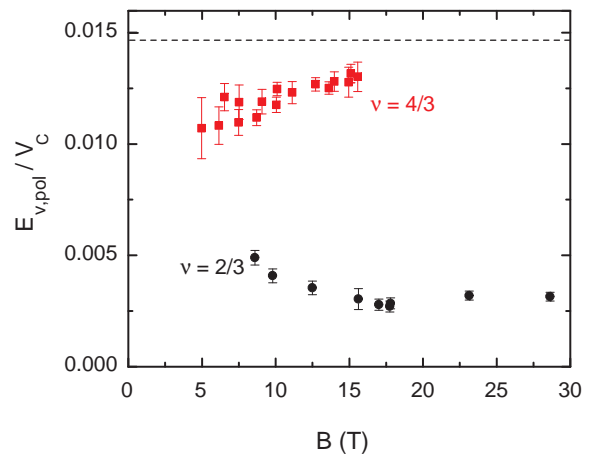}
\caption{Comparison of measured valley polarization energies for
$\nu = 2/3$ and 4/3, along with the theoretical prediction (dashed
line).}
\end{figure}

In Fig.\ 5, we summarize our main results by comparing the valley
polarization energies for $\nu = 2/3$ and 4/3. The data points
obtained from Fig.\ 4 are shown for $\nu = 2/3$, and the $\nu = 4/3$
data are obtained from similar measurements, including those
reported in Ref. \cite{padmanabhanPRB09}. Consistent with Ref.
\cite{padmanabhanPRB09}, we find a slight decrease in the value of
$E_{v,pol}$ with decreasing $n$ for the case of $\nu = 4/3$. $\nu =
2/3$ however, shows a slight increase with decreasing $n$. The main
observation in this figure is the glaring discrepancy between the
values of $E_{v,pol}$ for $\nu = 2/3$ and 4/3 thus implying a
breaking of the particle-hole symmetry. Note that comparison between
the two fractions is done at the same values of $B$, thereby
eliminating the influence of $B$-dependent parameters such as CF
effective mass.

In the past, there have been measurements of $spin$ polarization
energies of CFs in (single-valley) GaAs 2DESs around both $\nu =
3/2$ and 1/2 using a variety of techniques. Most of these studies,
however, focus on fractions around either $\nu = 3/2$ or $\nu =
1/2$. The reported values for $E_{Z,pol}/V_{C}$ (the CF
$spin$-polarization energy normalized to $V_{C}$) for $\nu = 2/3$
range from 0.009 \cite{engelPRB92,kukushkinPRL99} to 0.018 \cite{sternPRB04} while
the values for $\nu = 4/3$ lie between 0.019 \cite{kukushkinPRL99}
and 0.022 \cite{duPRB97}. Optical studies in Ref.\
\cite{kukushkinPRL99} indicate the spin-polarization energies for
both $\nu = 2/3$ and 4/3 as 0.009$V_{C}$ and 0.019$V_{C}$ (measured
at $B\simeq$ 2.5 T and 11 T respectively). A compilation of these results suggest
that the spin-polarization energies for $\nu = 4/3$ are on the
average higher than for $\nu = 2/3$ in spin-systems also. The
importance of particle-hole symmetry breaking, however, was not
discussed by any of the above studies.

Although higher values of polarization energy for $\nu = 4/3$
(compared to 2/3) are observed in spin systems also, the difference
is much more pronounced in our two-valley system (a factor of 4.3
for $B\simeq15$ T). This is possibly because of LL mixing. It is
known that LL mixing, parameterized by the ratio of Coulomb to the
cyclotron energy ($\hbar \omega_{c}$) of electrons, breaks particle
hole-symmetry \cite{macdonaldPRB84}. In AlAs 2DES, LL mixing is
especially significant due to the large band mass ($m_{b} = 0.46
m_{e}$ compared to $m_{b} = 0.067 m_{e}$ for GaAs; $m_{e}$ is the
free electron mass). For example, at a field of $B = 15$ T, the
values of $V_{C}$ and $\hbar \omega_{c}$ for AlAs are 252 K and 44
K, respectively. For the same $B$, the corresponding values for GaAs
are 194 K and 300 K. It is clear that AlAs is more prone to the
effects of LL mixing. Another unique feature of the AlAs 2DES is the
anisotropy of the Fermi contours at $B = 0$, the consequences of
which are unknown.

In Ref.\ \cite{parkPRL98} a theory was developed for the
spin-polarization energy of single-valley, two-spin systems. The
theory makes no distinction between the FQH states at $\nu = 2/3$
and $\nu = 4/3$ and predicts a $spin$ polarization energy of
0.0146$V_{C}$ (based on the polarization mass model
\cite{parkPRL98,padmanabhanPRB09}). This value is shown in Fig. 5 as
the dashed line. Our experimental results for $valley$ polarization
energies for $\nu = 2/3$ and 4/3 are both lower than this value,
with 4/3 showing a better agreement. The valley polarization
energies are also overall smaller than their experimentally measured
spin counterparts (in GaAs 2DESs), which show an overall better
agreement with the theoretical estimate. The relatively better
agreement between measured spin polarization energies and the theory
is consistent with the LL mixing being much smaller in GaAs systems.
Anisotropic electron-electron interaction, a result of elliptical
Fermi contours in AlAs 2DESs, might also bring about a difference
between the spin and valley degrees of freedom.

Independent of the theoretical understanding, we emphasize that the
data presented here highlight the breaking of the particle-hole
symmetry in our AlAs 2DES. Influence of LL mixing and Fermi contour
anisotropy needs to be studied further for a quantitative
understanding of this phenomenon.

We thank the NSF and DOE for financial support, and J.K. Jain for
illuminating discussions. A portion of this work was performed at
the National High Magnetic Field Laboratory, which is supported by
NSF Cooperative Agreement No. DMR-0654118, by the State of Florida,
and by the DOE. We thank G. Jones, S. A. Maier, T. Murphy, E. Palm
and J.H. Park for assistance.


\begin{references}
\bibitem{footnote2}
Ideally we expect the gap at $\epsilon = 0$ to be equal to the gap
at extreme values of $\epsilon$. The much smaller value of the gap
observed at $\epsilon = 0$ is likely linked to the proximity of the
two finite-width peaks in the piezoresistance trace.
\end{references}
\end{document}